

\hfill ULB-TH 8/92
\line{\hfill TAUP 2017-92}

\hfill December 1992
\bigskip
\bigskip

\title     {{\bf{BLACK HOLE TUNNELING ENTROPY AND THE SPECTRUM OF
GRAVITY}}\foot{Supported in part by NATO CRG 890404}}
\bigskip

\centerline{{\bf{A. Casher and F. Englert}}\foot{Postal address:
 Universit\'e
Libre de
Bruxelles, Campus Plaine, C.P.225, Boulevard du Triomphe, B-1050 Bruxelles,
Belgium.}}  \centerline{\it{ Service de Physique th\'eorique, Universit\'e
Libre
de Bruxelles, Belgium}}   \centerline{\it{and}}
\centerline{\it School of Physics and Astronomy}
\centerline{\it Raymond and Beverly Sackler Faculty of Exact Sciences}
\centerline{\it Tel-Aviv University, Ramat-Aviv, 69978 Tel-Aviv, Israel}

\bigskip
\smallskip

\centerline{ABSTRACT}
\smallskip

The tunneling approach for entropy generation in quantum gravity is
applied to black holes. The area entropy is recovered and shown to count
only a
tiny fraction of the  black hole degeneracy. The latter  stems  from the
extension of the wave function outside the barrier. In fact the
semi-classical
analysis leads to infinite degeneracy. Evaporating black
holes leave then infinitely degenerate  ``planckons" remnants  which can
neither
decay into, nor be formed from, ordinary matter in a finite time. Quantum
gravity opens  up at the Planck scale into an infinite Hilbert space which
is
expected to provide the ultraviolet cutoff required to render the theory
finite
in the sector of large scale physics. \vfil \eject

\chapter{{\bf{Introduction}}}

Tunneling in quantum gravity can generate entropy$^{[1],[2]}$. To
understand how
such an apparent violation of unitarity may arise, let us first
consider a classical space-time background geometry with compact Cauchy
hypersurfaces. If quantum fluctuations of the background are taken into
account,
quantum gravity leaves no ``external"  time parameter to describe the
evolution
of matter configurations in this background. Indeed, the solutions of the
Wheeler-De Witt equation$^{[3]}$  $${\cal H} \vert \Psi \rangle = 0 \eqno
(1) $$
where $\cal H$ is the Hamiltonian density  of the interacting
gravity-matter
system can contain no reference to such time when there are  no
contribution to
the energy from surface terms at spatial infinity. This is due to the
vanishing
of the time displacement generator and even though the theory can be
unambiguously formulated only at the semi-classical level, such a
consequence of
reparametrization invariance should have a more general range of validity.

To parametrize evolution, one then needs a ``clock" which could correlate
matter
configurations to ordered sequences of spatial geometries.  If quantum
fluctuations of the metric field can be neglected, the field components
$g_{ij}$ at every point of space can always be parametrized by a classical
time
parameter, in accordance with the classical equations of motion. This
classical
time, which is in fact a function of the $g_{ij}$, can  be used to
describe the evolution of matter and constitutes thus such a dynamical
``clock"
correlating matter to the gravitational field$^{[4]}$. This description is
available in the semi-classical limit of (1) where the classical background
  evolving in time is represented by a coherent superposition of W.K.B.
``forward" waves  formed from eigenstates of (1). When quantum metric field
fluctuations are taken into account, ``backward" waves, which   can  be
interpreted as flowing backwards in time, are unavoidably generated from
(1)
and the operational significance of the metric clock gets lost outside the
domain of validity of the semi-classical approximation.
Nevertheless, in domains of metric field configurations where both forward
and
backward waves are present but where quantum fluctuations are sufficiently
small, interferences with such ``time reversed" semi-classical solutions
will in
general be negligible\foot{For a recent discussion of related problems see
reference $[5]$.}. Projecting then out the backward waves restores the
operational significance of the metric clock but the evolution marked by
the
correlation time is no more unitary: information has been lost in
projecting
these backward waves stemming from regions where quantum fluctuations of the
clock are significant. This is only an apparent violation of unitarity
which
would be disposed of if the full  content of the theory would be
kept, perhaps eventually by reinterpreting backward waves in terms of
the creation
of ``universe" quanta through a further quantization  of the wavefunction
(1).

This apparent violation of unitarity is particularly marked if the
gravitational
clock experiences the strong quantum fluctuations arising from a tunneling
process. This can be illustrated from the simple analogy, represented in
Fig.1,
offered by a nonrelativistic closed system of total fixed energy $E$ where
a
particle in one space dimension plays the role of a clock for surrounding
matter and tunnels through a large potential barrier. Outside the barrier,
the
clock is well approximated  by semi-classical waves, but if on the left of
the
turning point one would take only forward waves, one would inevitably have
on
the right of the other turning point both forward and backward waves with
large
amplitudes compared with the original ones. The ratio between the squares
of the
forward amplitudes on the right and on the left of the barrier for a
component
of the clock wave with given clock energy $E_c$ is the inverse transmission
coefficient $N_0(E_c)$ through the barrier and provides a measure of the
apparent violation of unitarity.

In reference [2], it was shown that for a class of de Sitter type of
space-times, which admit compact Cauchy hypersurfaces, tunneling could
occur
between a ``wormhole" and an expanding universe. It was proven that,
for sufficiently small cosmological constant,  $\ln N_0$ played the role of
a
true thermodynamic entropy for the metric clock transferable to matter in
reversible processes. Explicit evaluation gave for this tunneling entropy
$\ln
N_0 = A/4$ where $A$ is the area of the event horizon. Thus one recovered
in this way the horizon thermodynamics of Gibbons and Hawking$^{[6]}$.

The tunneling entropy $\ln N_0$ is in last analysis an effect
of quantum fluctuations in quantum gravity. Therefore, despite the fact
that
no violation of unitarity would  appear in a complete description including
backward waves, this entropy should be expressible in terms of density of
states of matter and gravity. Tunneling offers an interesting perspective
in this
direction because it enlarges the semi-classical wave function of
space-time to
include in its description the other side of the barrier. Unfortunately,
for
the space-times considered above, the other side is a wormhole, that is in
the
classical limit simply a point on the Euclidean section of the original
manifold. This makes it illusory to describe in semi-classical terms the
configurations of the wormhole side of the barrier\foot{One could deform
the
Euclidean section to open the point into a Planckian size universe but this
would not change the conclusion.}.

In the present paper, we shall show that for black hole geometries, the
area
entropy is also interpretable in terms of tunneling. Now, Cauchy
hypersurfaces are not compact and evolution can be described by the
Minkowskian time available from surface terms at spacelike infinity.
However for spherically symmetric solutions, as long as no mass is brought
into
the system from infinity, the Minkowskian time appears as an irrelevant
unmeasurable phase in the quantum state of the system and the above
argument can be repeated. Tunneling amplitudes between spherically
symmetric
configurations with same total mass $M$ can still be searched for and the
related tunneling entropy is still expected to be  $\ln N_0$ and equal to
$A/4$.
This will indeed turn out to be the case but the crucial new feature which
will
emerge is that the black hole is not connected by tunneling to a wormhole
(in
fact, there are no such tunneling amplitudes), but to a large scale
matter-gravity configuration  describable in classical terms. These
configurations are analogous to macroscopic collapsing states frozen just
outside the Schwartzshild radius. We shall call these
configurations ``achronons" because they are, for the outside observer
deprived
of any time dependent properties as a consequence of an infinite time
dilation.
Achronons of given mass will be shown to be quantum mechanically
infinitely degenerate. Hence   black holes in quantum gravity have also
infinite degeneracy as their wave function is connected by
tunneling to the achronon side of the barrier. This means of course that
the
number of states $ \exp A/4 $ counted by the tunneling entropy is only a
finite
number of ``surface" states out of an infinite set of internal states which
cannot belong to the same finite Hilbert space as the matter surrounding
the
black hole in a finite volume. This mismatch will entirely modify the black
hole
evaporation process at its last stage. In fact the evaporation
must stop when the black hole   reaches the Planck scale, leaving a
stable ``planckon" remnant which can  neither be
created out of, nor decay into, ordinary matter in a finite external time.
Such objects were introduced previously$^{[7]}$ to avoid the violation of
unitarity which would arise from a complete black hole evaporation\foot{For
a
comprehensive review on recent attempts to solve the black hole unitarity
puzzle, see reference ${[8]}$.}. They follow here  directly from   the
tunneling
structure of the black hole-achronon wave function.

To strengthen these tentative conclusions, one should improve the present
analysis in two respects. First, we have been restricted by the
semi-classical treatment of quantum gravity and we can only surmise that
the
crucial element which came out of it, namely the infinite degeneracy of the
black hole wave function, will survive the full quantum description.
Second,
although the semi-classical treatment defines unambiguously the achronon
from
the tunneling, we have not realised such a configuration in a genuine field
theoretic way. We have only examplified its features by a phenomenological
model, which although consistent, is too schematic to be directly
physically
relevant. Hopefully a more complete and realistic illustration of the
achronon
will secure the explicit construction of the black hole-achronon wave
function.

Notwithstanding these limitations, the present approach gives strong
support to
the planckon hypothesis. It would of course be of great interest to find at
least some indirect evidence in favour of their existence. This is not
totally
impossible as planckons may have interesting cosmological and astrophysical
consequences if they where present in the early universe as might be the
case if
primordial black holes played an important role in cosmogenesis$^{[9]}$. At
a
more fundamental level, they would provide, as a consequence of unitarity,
a
natural cut-off at the Planck size for the ultraviolet spectrum of Hilbert
space
of states describing large scale physics and are therefore expected to
render
quantum gravity expressible as a finite theory.

In the presentation of the paper, rather than deducing achronon
configurations from the analysis of black hole tunneling amplitudes, we
found it
more convenient to motivate the latter by first introducing achronons as
classical solutions of general relativity. This is exemplified in section 2
in a simple shell model. In section 3, we
show how achronons surrounding a black hole can screen their temperature to
zero
or to a finite quantity if the black hole has, classically, a vanishingly
small
Schwartzschild radius. These properties are then used in section 4 to prove
that
eternal black hole are related by tunneling to achronon configurations. The
inverse transmission coefficient is computed in the semi-classical limit
and its
relation to entropy is proven. Contact is made  between the black hole
tunneling
entropy and the Gibbons-Hawking thermodynamics. In section 5, the
infinite degeneracy of quantum black holes of given mass is established.
The
nature of the planckon remnants follows then from the evolution of the
potential
barrier during the black hole decay. Their properties are reviewed and
their
bearing on the spectrum and the scope of quantum gravity is discussed.
Mathematical details are relegated to the Appendix.

\chapter{{\bf{ The Achronon}}}


Our basic action in four dimensional Minkowski space-time will be
$$S=S_{grav}+S_{matter}\eqno(2)$$
where $ S_{grav}$ has the conventional form ($G=1$) :
$$ S_{grav}=-{1\over 16\pi} \int \sqrt{-g}R\, d^4x \eqno (3)$$
and $S_{matter}$ contains sufficiently many free parameters  to allow for
the
stress tensors considered below. In describing classical solutions, it
should be
kept in mind that they have to be interpreted as semi-classical solutions
of
(1). In particular, when using shells of infinite
energy density, eventual smearing out by quantum spread should be
understood.

Spherically symmetric solutions in general relativity are entirely
determined in terms of the energy density function $\sigma = T^0_0$ and the
radial pressure function $p_1 = -T^1_1$  in the coordinate system
$$ ds^2 = g_{00}(r)\,dt^2 - g_{11}(r)\, dr^2 - r^2(d\theta^2 +
\sin^2\theta \, d\phi^2) \eqno (4) $$
where we have restricted ourselves to
$t$independent configurations. The metric tensor is given by
 $$\eqalign{r_2[g^{11}(r_2) - 1] -r_1[g^{11}(r_1) -1] &=-2\int_{r_1}^{r_2}
dM(r)
\cr dM(r)&= 4\pi \sigma r^2 dr \cr} \eqno(5)$$ and
$${g_{00}(r_2)g_{11}(r_2)\over g_{00}(r_1)g_{11}(r_1)} =\exp
\int_{r_1}^{r_2}8\pi r (\sigma + p_1) g_{11}\, dr. \eqno(6) $$
 For asymptotically
flat solutions one chooses $g_{00} =1$ at $\infty$; in absence of black
hole
horizon, $ g_{11}>0$ and the solution considered is everywhere static. The
other
components of the (diagonal) energy-momentum tensor $p_{\theta} = -
T_{\theta}^{\theta}$ and $p_{\phi}=-T_ {\phi}^{\phi}$ are determined by the
Bianchi identities and can be expressed in terms of $p_1$ and $\sigma$:
$$p_{\theta}= p_{\phi}= {1\over4}(\sigma + p_1) {8\pi r^2 p_1
+ 2M(r)/r \over 1-2M(r)/r}+{1\over2}rp^{\prime}_1+p_1. \eqno(7)$$

Let us now consider a static spherically symmetric distribution of matter
surrounded by an extended shell comprised between two radii $r_a$ and
$r_b$. We define
$$ \hat\sigma
\equiv \int_{r_a}^{r_b}\sigma g_{11}^{1/2}dr, \quad \hat p_\theta \equiv
\int_{r_a}^{r_b}p_\theta g_{11}^{1/2}dr, \quad \hat p_1 \equiv
\int_{r_a}^{r_b}p_1 g_{11}^{1/2}dr. \eqno(8)$$
Assuming $p_1=0$, one may perform the thin shell limit  $r_b \to r_a =R$ in
these integrals by using  $dM(r) =
 4\pi \sigma R^2 dr$. From (5) and (7) one then gets
$$ \eqalignno {4\pi R \hat\sigma &=(1-2m^-/R)^{1\over2} -
(1-2m/R)^{1\over2}
&(9)\cr 8\pi R \hat p_\theta & = {1-m/R \over (1-2m/R)^{1\over2}} -
{1-m^-/R
\over (1-2m^-/R)^{1\over2}} &(10)\cr \hat p_1 &=0 &(11)\cr } $$
where $m$ and $m^-$ are the values of $M(r)$ respectively at $r_b$ and
$r_a$
and $m_s = m - m^-$ is thus the mass of the shell. Equations (9) and (10)
are
the standard result$^{[10]}$. As the radius $R$ approaches $2m$, these
solutions become physically meaningless when  $\hat p_\theta$  becomes
greater than $\hat\sigma$; this violates indeed the  ``dominant energy
condition"$^{[11]}$, implying the existence of observers for which the
momentum
flow of the classical matter becomes spacelike. In fact, the shell is
mechanically unstable even before this condition is violated$^{[12]}$.

The divergence of $\hat p_\theta$ when
$R \to 2m$  appears in (10) because of the vanishing denominator in
(7).   Equation (10) depends however crucially on the radial pressure being
zero inside the shell. Relaxing this condition we see indeed that  a finite
value of $p_1$ multiplies in (7) the energy density $\sigma$ which becomes
infinite in the thin shell limit. It is in fact possible to avoid all
singularities of the stress tensor as $R \to 2m$ by requiring $p_1$  inside
the
shell to satisfy, before performing the thin shell limit,   $$ 4\pi r^2 p_1
+
{M(r)\over r}=0 .\eqno(12)$$ Inserting the solution of (12) back in (7), we
get
for the trace of the energy momentum tensor $T^\mu_\mu$ the equation of
state
 $$ T^\mu_\mu =2\sigma \eqno(13)$$
which means that the source of the
``Newtonian" force $T^0_0 -1/2 \delta^0_0 T^\mu_\mu$  due to the any inner
part of the shell on the remainder vanishes.  An alternate
way to discover the solution (12) is precisely to impose the trace
condition
(13) in equation (7): the solution of this differential equation with
$p_1(r_a)$ fixed (and equal to $ -M(r^a)/4\pi (r^a)^3$) is equation (12).

This solution is  unsatisfactory if the (extended) shell sits
in an arbitrary background because of the finite discontinuity of the
radial
pressure across the shell boundaries which would lead to singularities in
$p_\theta$. We may ensure continuity of the radial pressure by immersing
the
shell in suitable left and right backgrounds. To avoid reintroducing stress
divergences when $r^b$ approaches $2M(r^b)$ these should satisfy $(\sigma
+
p_1)=0$ at the shell boundaries. One can now perform the thin shell limit.
The
finite discontinuity of $p_1(r)$ at $r =R$ leads to
$$  \hat p_\theta = - \hat {\sigma \over 2},\quad
\hat p_1 =0\eqno(14) $$
instead of (10),(11) and $ \hat \sigma$ is still given by
(9).  The dominant energy condition is satisfied everywhere, as is the
``weak
energy condition"$^{[11]}$ ensuring positivity of the energy density for
any
observer. Provided the background is smooth enough in the neighbourhood of
the
shell, no stress divergences will appear when it approaches the
Schwartzshild radius.

Such static thin shells sitting outside the Schwartzschild
radius but infinitesimally close to it will be referred to as ``limiting
shells". The mass $m_s$ of the limiting shell plus the mass $m^-$ of  the
inner
matter contribution is  equal to the black hole mass whose horizon would be
at
the limiting radius $r=2m$. The striking feature of the region bounded by
the
limiting shell is that it gives rise to an infinitely large time dilation
in the
global Schwartzshild time. Indeed
$$ g_{00}(r)=(1-{2M(r)\over r})
\exp{\left[- \int_r^{\infty}8\pi r^\prime(\sigma + p_1)  g_{11}\,
dr^\prime\right]}\eqno(15) $$
and performing the explicit integration over the
shell, we get in the region
$0 \leq r<2m$
$$ g_{00}(r)=(1-{2M(r)\over r})\left[{R-2m \over R-2m^-}\right]
\exp{\left[-{\cal R}\int_r^{\infty}8\pi r^\prime(\sigma + p_1)  g_{11}\,
dr^\prime\right]}.\eqno(16) $$
Here the radius $R$ of the shell is taken at $R= 2m + \epsilon$ where
$\epsilon$ is a positive infinitesimal and the symbol $\cal R$ means that
the
integral is carried over the regular matter contribution only. Clearly, $
g_{00}(r) = O(\epsilon)$ for $0 \leq r<2m$, $t$ arbitrary.

This domain of space-time is characterized  by a Killing vector which
is light-like in the limit $\epsilon \to 0$. When the space-time geometry
presents a 4-domain endowed with such a limiting
light-like Killing vector, we shall call the domain an achronon. All
spherically symmetric achronon configurations will exhibit  an infinite
dilation
of the Schwartzshild time $t$  with respect to the outside world, or
equivalently, massless modes emitted by the achronon are infinitely
redschifted.
Classically, the achronon has  the ``frozen"
appearance  of a collapse at infinite Schwartzshild time. The difference
is
that it is  also frozen  in  space-time. This is the reason
why, in contradistinction to collapsing shells,  stresses (14) were
needed to build the purely static solution  considered above. However,
requiring
exact staticity everywhere in space-time is mathematically convenient but
perhaps a too stringent and physically unnecessary constraint. Thus our
shell solution (even if extended to a finite width) and the concomitant
restriction on the background, should be viewed as a simple illustrative
model.
More elaborate achronon solutions
will be discussed elsewhere.

\chapter{{\bf{  Thermal Screening of a Black Hole}}   }

Up to now, we have considered achronons in a trivial space-time topology
but
they can also be introduced in the topology of an eternal black hole. An
eternal
black hole of mass $m_0$ contains two asymptotically flat Schwartzshild
patches
connected by a throat. It represents the maximal extension of
the Schwartzschild solution which is singular at $r=0$ and  is dynamical
outside
the patches as  seen from the well known Kruskal representation (Fig.2).
One can
add in the patches a static distribution of matter without
changing the topology as long as  outer horizons are avoided. We shall
consider such distributions and we shall limit ourselves to matter
configurations which are identical in both patches. Thus, achronons of mass
$m -
m_0$ surrounding a black hole of mass $m_0$ are defined in this topology by
their matter distribution in a static patch.

Let us now consider such a achronon, possibly surrounded by static matter.
Using
the metric (4) in the static patches, $M(r)$ is defined in general  from
(5) for
$r>2m_0$ by $$ M(r)= m_0 + \int^r_{2m_0} 4\pi \sigma r^{\prime
2}\,dr^{\prime}.
\eqno(17) $$ The Kruskal metric is $$ ds^2={dT^2-dX^2\over F^{\prime
2}(\xi)} -
r^2(\xi)\,(d\theta^2+\sin^2 \theta\, d\phi^2) \eqno (18) $$
which   is related to the static metric (4) within a patch by
$$ g_{11}^{1/2}\,dr=d\xi, \quad \xi =0\  \hbox{at the horizon}, \eqno(19)
$$
and
$$ F(\xi)=\sqrt{X^2-T^2}. \eqno(20) $$
A Cauchy hypersurface $\Sigma_c$
represented in Kruskal coordinates by $T=0$ (Fig.2) connects the space-time
with Minkowskian signature to a solution of the Euclidean Einstein
equations.
The latter can be described by the metric (4) with $t=-it_e$  and is
periodic in
the Euclidean time $t_e$.  The
Euclidean period ${\cal T}^{-1}$ can then be computed from the metric (4)
in the
vicinity of the black hole horizon $r_0$: $${\cal T} = {1\over
4\pi}[g_{00}(r_0)\,g_{11}(r_0)]^{-1/2} {dg_{00}(r)\over dr}\vert_{r=r_0}
\eqno(21)$$ and from (5) and (6) one gets $$ {\cal T}= {1\over 8\pi m_0}
\exp{\left[-\int_{2m_0}^{\infty}{4\pi r^\prime (\sigma + p_1)  g_{11}\,
dr^\prime}\right]}. \eqno(22) $$
Comparing (22) with (15), one immediately sees
that the inverse Euclidean period of a black hole surrounded by an achronon
vanishes.

To illustrate this phenomenon consider an achronon of mass $m-m_0$ bounded
by a
limiting shell of mass $ m_s \leq m-m_0$. The limiting shell sits at a
radius
$R = 2m + \epsilon$ and we may rewrite (22) as  $$ \eqalign {{\cal
T}={1\over
8\pi m_0}\lim_{\epsilon \to
0}&\exp{\left[-\int_{R-\epsilon}^{R+\epsilon}4\pi
r^\prime(\sigma + p_1)  g_{11}\, dr^\prime \right]}\cr
&\exp{\left[-\int_{2m_0}^{R-\epsilon}4\pi r^\prime(\sigma + p_1)  g_{11}\,
dr^\prime\right]}  \exp{\left[-\int_{R+\epsilon}^{\infty}4\pi
r^\prime(\sigma +
p_1)  g_{11}\, dr^\prime \right]}.\cr} \eqno(23)$$
 The first factor
is easily evaluated in the limit $\epsilon \to 0$ and (23) yields   $$
{\cal
T}={1\over 8\pi m_0}\left[{R-2m\over R-2m^-}\right]^{1\over2}
\exp{\left[-{\cal
R}\int_{2m_0}^{\infty}4\pi r^\prime(\sigma + p_1)  g_{11}\,
dr^\prime\right]}\eqno(24)$$
where $ m^- = m - m_s$. A glance at (24) shows that the limiting inverse
Euclidean period when $\epsilon \to 0$ is indeed $${\cal T}_{\epsilon \to
0}=0
\eqno(25) $$

We know, from the work of Gibbons and Hawking$^{[7]}$ that ${\cal T}$ in
(21)
is the temperature at infinity  of quantum matter in the background
of the classical  gravity-matter system considered and is its
equilibrium temperature in the energy conjugate
to the static time $t$. In particular, when no matter surrounds the black
hole,
${\cal T}$ reduces to the usual black hole temperature $1/ 8\pi m_0$. We
shall show in the following section that ${\cal T}$ is also the equilibrium
temperature, in the semi-classical limit of quantum gravity, of the
interacting
gravity-matter system itself. Thus ,(25) implies that the thermal effects
of a
black hole of mass $m_0$ can be entirely screened by a achronon of mass
$m-m_0$, as expected from the infinite redshift due to the achronon.

A different situation can however arise if the black hole has, classically,
a
vanishing small mass. Such an object, which we shall call a germ black
hole,
generates a non trivial topology. As long as the surrounding matter does
not
form a achronon,  the value of ${\cal T}$ tends to infinity when the mass
$m_0$
of the germ tends to zero. But in the
presence of an achronon the resulting
Euclidean periodicity can take any value, depending on the limiting
process. In
particular, one may have
 $$ {\cal T}_{\epsilon \to 0} = {\cal T}_m \eqno (26)$$
where ${\cal T}_s$ is the inverse Euclidean period of a black hole of mass
$m$
surrounded by the same matter distribution as the corresponding achronon of
mass $m-m_0$. This is exemplified in (24) by letting $R-2m$ go to $0$  as
$Cm_0^2$ and tuning the constant $C$ to satisfy (26).

The possibility of constructing, in presence of a germ, an achronon
with the same behaviour in Euclidian time as a genuine black hole will be
the
key to the tunneling between achronons and black holes.

\chapter{{\bf{The Black Hole Tunneling Entropy}}}

We first review\foot{For a more detailed discussion see reference [2].}
and generalize to the present case the
description of tunneling in quantum gravity, obtained in reference [2] for
geometries with de Sitter topology.  Consider in general two spacelike
hypersurfaces $\Sigma_1$ and $\Sigma_2$ which are turning points in
superspace
(or turning  hypersurfaces) along which   solutions  of the Minkowskian
classical equations of motion for gravity and matter meet a classical
solution
of their Euclidean extension. $\Sigma_1$ and $\Sigma_2$ are thus the
boundaries
of a   region $\cal E$ of  Euclidean space-time defined by the Euclidean
solution.  If $\cal E$ can be continuously shrunk to zero one can span
$\cal E$
by a continuous set of hypersurfaces $\tau=$ constant such that $\tau
\equiv
\tau_1$ on  $\Sigma_1$ and $\tau \equiv \tau_2$ on  $\Sigma_2$. These
$\tau=$
constant surfaces define a coordinate system which we shall call
synchronous; the
Euclidean metric in $\cal E$ can be written in the form $$ds^2 =
N^2(\tau,x_k)\,
d \tau^2+g_{ij}(\tau,x_k)\,dx^i\, dx^j \eqno(27)$$ where $N(\tau,x_k)$ is a
lapse
function. The Euclidean  action $S_e$ over $\cal E$, from $\Sigma_1$ to
$\Sigma_2$, is obtained by analytic continuation from the Minkowskian
action
(2) and can be written as
 $$\eqalign{S_e(\Sigma_2,\Sigma_1)   =\int_{\cal E}\Pi^{ij}\partial_\tau
g_{ij}\,d^4x +\int_{\cal E} \Pi^a\partial_\tau \phi_a\, d^4x &-  \int_{\cal
E} {
d\over d\tau } (g_{ij} \Pi^{ij}) \,d^4x\cr  &-{1\over 8\pi}\int_{\cal E}
\partial_k[(\partial_j N)g^{kj}\sqrt {g^{(3)}}]\,d^4x. \cr}  \eqno (28) $$
 Here $\Pi^{ij}$ and $\Pi^a$ are the Euclidean momenta conjugate to the
gravitational fields $g_{ij}$ and to the matter fields  $\phi_a$; $g^{(3)}$
is
the three dimensional determinant.

On the turning hypersurfaces $\Sigma_1$ and $\Sigma_2$, all field momenta
$(\Pi^{ij},\Pi^a)$ are zero in the synchronous system and the third term in
(28)
vanishes.  The last term in (28) also vanishes if the
hypersurfaces $\Sigma_1$ and $\Sigma_2$ are compact (which was the case
considered in reference $[2]$) but may receive contributions from infinity
otherwise. We shall have to consider here the case where the two non
compact
turning hypersurfaces merge at infinity  so that
the Euclidean action $S_e(\Sigma_1,\Sigma_2)$ does not get contributions in
$\cal E$ from the last term in (28).  The classical Minkowskian solution in
the
space-time ${\cal M}_1$ containing $\Sigma_1$ can be represented quantum
mechanically by a ``forward wave" solution $\Psi(g_{ij},\phi_a)$ of the
Wheeler-de Witt equation (1) in the semi-classical limit. At $\Sigma_1$,
this
wave function enters, in the WKB limit, the Euclidean region ${\cal E}$ and
leaves it at $\Sigma_2$ to penetrate a new Minkowskian space-time ${\cal
M}_2$.
The tunneling of $\Psi(g_{ij},\phi_a)$ through  ${\cal E}$ engenders in
addition
to the  ``forward wave" solution a time reversed ``backward wave". The
inverse   transmission coefficient $N_0$ through the barrier  measures the
ratio
of the norms of the forward waves at  $\Sigma_2$ and  $\Sigma_1$. For large
$N_0$ one may write in the synchronous system $$ N_0 = \exp {-[2(\int_{\cal
E}\Pi^{ij}  \partial_\tau g_{ij}\,d^4x +\int_{\cal E} \Pi^a \partial_{\tau
}\phi_a\, d^4x)]}. \eqno (29) $$ As all surface terms in (28) vanish in this
 system, (29) can be rewritten in the coordinate invariant form $$
N_0 = \exp {[2S_e(\Sigma_1,\Sigma_2)]}. \eqno (30) $$

Consider now an eternal black hole of mass $m$ surrounded by a spherically
symmetric distributions of matter, the same in both static patches. Compare
this
classical solution of general relativity to another one consisting of an
achronon
of mass $m-m_0(\epsilon)$  surrounded by the same matter
distribution and screening a germ black hole of mass $m_0(\epsilon) \to 0$
to
the same Euclidean period. Both solutions are thus characterized by the
same
total mass $M$, the same matter distribution of mass $M-m$ outside the
radius
$2m +\eta$, $\eta$ infinitesimal\foot{For the shell model of section 2, one
may
take $\eta = \epsilon$ as $g_{00}$ and $R-2m$ are of the same order of
magnitude. For sake of generality we do not impose this relation here.} and
the
same Euclidean period ${\cal T}^{-1}$. We shall identify  ${\cal M}_1$ with
the
achronon solution and ${\cal M}_2$ with the black hole one. We label  by
$\Sigma_c^{B.H.}$ and $\Sigma_c^A$ respectively the turning hypersurfaces
in
the black hole and in the achronon geometries.

$\Sigma_c^{B.H.}$ and $\Sigma_c^A$ can be represented in Kruskal
coordinates by hypersurfaces $T=0$ and are depicted in Fig.3. They belong
to
Euclidean sections of these solutions ${\cal E}^{B.H.}$ and ${\cal E}^A$
which
can be described by Euclidean Kruskal metrics (17) with Euclidean time $T_e
=
iT$ or by static coordinates (4) with a periodic Euclidean time $t_e =it$;
for
both solutions the period has the same value ${\cal T}^{-1}$. It is clear,
from
the static coordinate description, that the two Euclidean space-time
geometries
${\cal E}^{B.H.}$  and ${\cal E}^A$ coincide for $r>2m + \eta$ but, while
the
Euclidean black hole terminates at $r=2m$, the achronon solution has an
extra
``needle" in the region $0< r<2m$ whose 4-volume is of order $\epsilon$.

We now identify, at finite $\eta$, $\Sigma_1$ with $\Sigma_c^A$ and
consider instead of a second turning hypersurface  $\Sigma_2$ a
hypersurface
$\Sigma_c^{\prime B.H.}$ which lies in ${\cal E}^{B.H.}$ and is such that
$r> 2m +\eta$ everywhere on it. $\Sigma_c^{\prime B.H.}$ is then contained
in the intersection of ${\cal E}^{B.H.}$ and of ${\cal E}^A$. When
$\eta \to 0$, $\Sigma_c^{\prime B.H.}$ can be taken arbitrarily close to
$\Sigma_c^{B.H.}$ and we shall prove in the Appendix  that all
gravitational
momenta on $\Sigma_c^{\prime B.H.}$ in a synchronous system vanish in this
limit.  We may then identify  $\Sigma_c^{\prime B.H.}$ with
$\Sigma_2$. The region $\cal E$ is thus contained in the needle
$0<r<2m+\eta$ of ${\cal E}^A$. Because of the Kruskal twofold symmetry
$\Sigma_c^A$ is mapped onto itself by a Euclidean time rotation of
half a period and thus $\cal E$ spans only half the needle 4-volume. From
(28),
we learn that the inverse transmission coefficient $N_0$  is simply the
exponential of the total Euclidean action of the needle. Although the
limiting
4-volume of the needle vanishes, the action is computable as the difference
between the Euclidean action of the black hole $S_e^{B.H.}$ over  ${\cal
E}^{B.H.}$ and of the achronon $S_e^{A}$ over ${\cal E}^A$. This difference
is
finite and well defined by cutting off the two spaces at an arbitrary
radius
$r_c$ greater than $2m$ as the two geometries and the two actions coincide
for
all $r >r_c$. We thus write $$N_0= \exp [S_e^{B.H.}-S_e^A]. \eqno(31)$$ To
evaluate these actions we take advantage of the covariance to express them
in
terms of the static coordinate system with gravitational and matter momenta
everywhere vanishing. Thus only the last surface integral in (28)
contributes
now to the action and can be expressed as
 $$S_e=
\int {d\over dr} \left[{-r^2\over
4 {\cal T}}[g_{00}g_{11}]^{-1/2} {dg_{00}\over dr}\right] \,dr. \eqno(32)$$
Using (21) and the fact that the integrand is the same at $r_c$ for
$S_e^{B.H.}$ and for $S_e^A$, we get
$$N_0= \exp {[4\pi m^2 - 4\pi m_0^2(\epsilon \to 0)]}  \eqno (33)$$
or, as $m_0$ vanishes in the limit,
$$N_0= \exp {A/4} \eqno (34) $$
where $A=16 \pi m^2$ is the area of the event horizon of the black hole.

We have thus learned that black holes are related by quantum tunneling
to another classical solution for gravity and matter, namely to an
achronon\foot{ One might have thought that a single point on ${\cal
E^{B.H.}}$
would represent, in the classical limit, a wormhole-like turning
hypersurface.
This is not the case as there is no continuous  mapping of
$\Sigma_c^{B.H.}$ onto
a point. More generally, one can show that no lower manifold contained in
${\cal E^{B.H.}}$ constitutes a turning hypersurface.}. Each achronon is
connected through a ``potential" barrier to a corresponding black hole of
mass
$m$. Let us tentatively take boundary conditions in field space by
assigning pure
forward  waves to achronons; the  relative probability of
finding a black hole with respect to an achronon is then $N_0$, since in
the
classical limit interferences between black holes propagating forward or
backward in time must be negligible.

Consider then two distinct achronons surrounded by matter distributions
such
that the total mass $M$ is the same for both classical solutions but $m$
and the
surrounding mass $M-m$ need  not be the same. Each achronon is related by
tunneling to its corresponding black hole and the ratio of inverse
transmission coefficients $N_0$ between the two achronon-black hole
configurations is, from (34), equal to  $\exp \Delta A/4$, where
$\Delta A$ is the change in black hole area. From the differential Killing
identity of reference [13], or equivalently from the variation of the
integrated
constraint equation over a static patch$^{[2]}$,we have
$$   \delta M -{\cal T} { \delta A \over 4 }=\delta_\lambda H_{matter}
\eqno(35)
$$ where $\lambda$ labels the {\it explicit\/} dependence of the matter
hamiltonian   $H_{matter}$ deduced from the action (2) on all (non
gravitational) ``external" parameters. As we are considering only
contributions
with fixed total mass $M$, it follows from (35)  that matter configurations
with neighbouring energies in a static patch of the black hole would be
Boltzmann
distributed at the global temperature ${\cal T}$  provided achronons with
different mass are taken to be equally probable. This is indeed a
consistent
assumption as all configurations describing achronons and surrounding
matter
with fixed total mass $M$ have, from (35) with $A=0$, the same total energy
and
may be described by a microcanonical ensemble. Thus the  temperature of the
static patch is indeed $\cal T$. Therefore (35) also implies that  $\delta
A/4$
is the differential entropy of the black hole and that the latter is in
thermal
equilibrium with the surrounding matter at the temperature $\cal T$. As the
entropy must be an intrinsic property of the black hole, not
only is equilibrium a consequence of the chosen boundary conditions in
field
space but the converse is also true: the temperature obtained directly from
(35)
with   entropy identified as $A/4$ must agree at equilibrium with the
thermal
distribution generated from the field boundary conditions. This justifies
a posteriori the above choice of boundary conditions\foot{up to changes
which
would not alter the probability ratios in the large $N_0$ limit}.

The tunneling approach to the horizon entropy and temperature$^{[1],[2]}$
applied  here to the black hole  differs from the analysis based on the
Euclidean
periodicity of Green's functions$^{[6]}$ in two respects. On the one hand,
the
present approach yields the thermal spectrum, and then the entropy, from
the
{\it backreaction\/}  of the thermal matter on the gravitational field,
in
contradistinction to the Green's function approach. On the other hand
however,
the thermal matter considered here is taken in the classical limit while
the
Green's function describes genuine quantum radiation. Both methods fall
short of a fully consistent quantum treatment of the backreaction problem
but
the  interpretation of the horizon entropy from tunneling will permit us to
uncover the quantum states building the black hole entropy;  in
fact, we shall see that the number of states $ \exp A/4$    count only a
minute fraction of the full black hole degeneracy.

\chapter{{\bf{ From Achronons to Planckons }}}

The entropy $A/4$ which can be exchanged reversibly from a black hole to
ordinary matter was rederived in the preceding section from the existence
of a
``potential barrier" between a black hole of mass $m$ and an achronon of
the
same mass. This was done in the context of an eternal black hole admitting
a
Kruskal twofold symmetry with two achronons separated by an
infinitesimal throat, each imbedded in the  surrounding geometry of a
static
space emerging from the eternal black hole throat. Within each space
black
hole-achronon states are in thermal equilibrium with their surroundings. We
are
therefore led to picture a black hole-achronon state , in the
semi-classical
limit, as a quantum superposition of two coherent (normalized) states,
$\vert
B.H. \rangle$ and $\vert A. \rangle$ representing respectively a classical
black
hole and a classical achronon. The
relative weight of the two states in thermal equilibrium is
approximately, up
to a phase, $\exp (-A/8)$ . It  follows from detailed balance at
equilibrium
between radiated matter and the black hole that the same superposition
should
hold for a the black hole who would only emit (and not receive) thermal
radiation at the equilibrium temperature. As a black hole formed from
collapse
indeed emits such a thermal flux, we infer that its state $\vert C \rangle$
should contain an achronon component with the same weight as in thermal
equilibrium. We thus write $$ \vert C \rangle = \vert B.H. \rangle + \exp
(-A/8)\vert A. \rangle. \eqno(36)$$

To a single black hole configuration one may associate many
distinct classical achronon configurations. In the shell model, for
instance,
there are infinitely many distinct classical matter configurations of the
same
total mass $m$. The argument is however much more general and   infinite
quantum degeneracy of the achronon is a direct consequence of the infinite
time
dilation. Indeed, the Hamiltonian $H$ is of the form  $$ H= \int
\sqrt{g_{00}}K(\phi_a , g_{ij}, \Pi_a, \Pi_{ij})\,d^3x \eqno(37)$$
and all its eigenvalues are squashed towards zero by the Schwartzschild
time
dilation factor $\sqrt{g_{00}}$, thus generating an infinite number of
orthogonal zero energy modes on top of the original achronon. In the
classical
limit, the phase space of zero-energy solutions  becomes infinite and  the
Killing identity (35) with $A=0$  confirms that the modes give no
contribution
to the achronon mass $m$. The same conclusions can be arrived at by
considering
the wave equation instead of the canonical Hamiltonian. For example, the
scalar
wave equation is $$ {1\over \sqrt g} \partial_\mu\sqrt g  g^{\mu
\nu} \partial_\nu \Phi =0. \eqno(38) $$
Thus the frequency of any mode is
proportional to $ \sqrt{g_{00}}$ and vanishes in the limit $g_{00}
\to 0$. By imposing boundary conditions at the surface,
$\phi$ may be expanded in creation and annihilation operators for the above
modes, thus realizing the degenerate spectrum.

The infinity of zero energy modes around  any background implies an
infinite
degeneracy of achronons of given mass and thus an
infinity of distinct quantum black hole states of the same mass differing
by
the achronon component of their wave function. This infinite degeneracy of
the
quantum black hole  provides  the reservoir from which are taken the finite
number of ``surface" quantum states $\exp A/4$ counted by the area entropy
$A/4$ transferable reversibly to outside matter.

Except for providing a rational for the large but finite testable entropy
of the
black hole, achronons do not modify the behaviour of large macroscopic
black
holes. However when their mass is reduced by evaporation and approaches the
Planck mass the barrier disappears and quantum superposition completely
mixes the
two components. Of course, this means that both the description in terms of
semiclassical configurations and of tunneling disappears. What remains however
as a consequence of unitarity, is the infinity of distinct orthogonal
quantum
states available which have no counterpart in the finite number of decayed
states. The quantum black hole  has become a planckon$^{[7]}$, that
is a planckian mass object with infinite degeneracy. Causality and
unitarity
prevent the decay (and the production) in a finite time of such
object$^{[7]}$,
and the argument applies to the ``parent achronon" as well. Indeed, if a
state
$\vert A_i \rangle $ of finite size and mass $m$ decays, or is produced,
within
a finite time $\tau$ in an approximately flat space-time, the total
number of
possible final states is limited by the number ${\cal N}$ of orthogonal
states
with total mass $m$ in a volume $\tau^3$. From unitarity, the degeneracy
$\nu
(m)$ of the  states  $ \vert A_i \rangle$ is at most ${\cal N}$. Thus if
$\nu (m)
\to \infty$, the time $\tau $ tends to infinity. Thus, achronons and in
particular planckons can neither decay nor be formed in a finite time.

As recalled in the introduction, planckons are a solution to the unitarity
puzzle \foot{see also reference[14].}
arising from black hole evaporation and may have played a crucial role in
seeding our universe and its large scale structure. At a more fundamental
level they have far reaching implications on the spectrum  of
quantum gravity. The opening at the Planck size of an infinite number of
states, an unavoidable consequence of the existence of planckons, may
appear as a
horrendous complication which could make quantum gravity definitely
unmanageable
but hopefully the converse may be true. Indeed planckons should make
quantum
gravity ultraviolet finite. The Hilbert space of physical states available
to
macroscopic observer must be orthogonal to the infinite set of states
describing
planckian bound states. Their wave function at planckian scales where
planckon
configurations are concentrated are therefore expected  to be vanishingly
small.
In this way, planckons would provide the required short distance cut-off
for a
consistent field theoretic description of quantum gravity within our
universe
while leaving the largest part of its information content hidden at the
Planck
scale.

An operational formulation of quantum gravity applicable within our
universe and
based on  conventional four dimensional gravity,  may thus well be within
reach. But it is nevertheless tempting to dwell upon the further
significance of
the picture that emerges. The sudden widening of the spectrum of physical
states
at the Planck scale strongly suggests that the relative scarcity of states
which describe large distance physics (as compared to the Planck
size) is
 due to the fact that the exsistence of
 observables  whose correlations survive at
 macroscopic range is contingent on the notions of scale and metric.
The appearance of these these concepts in the organization of long
long distance physics
                                     , at the cost of relegating most of
the
information to the Planck scale ,
                             would imply that scale should  be absent from
a
fundamental description of the physical world. Consistency may then
ultimately
require a unified theory, of which string theory is perhaps a precursor,
which
by eliminating the gravitational scale from the basic formulation would
render
obsolete the use of  a standard of length or of time.
\bigskip
\centerline{{\bf{Acknowledgements}}}

We are very grateful to R. Balbinot, R. Brout,J. Katz, J. Orloff, R.
Parentani
and Ph. Spindel for most enjoyable, stimulating and clarifying discussions.
\endpage


\appendix
In computing the tunneling amplitude from the achronon turning
hypersurface  $\Sigma_c^A$ to the black hole hypersurface
$\Sigma_c^{B.H.}$,
we have, in section 4, replaced $\Sigma_c^{B.H.}$ by a hypersurface
$\Sigma_c^{\prime B.H.}$ which lay in the intersection of the Euclidean
sections
of the achronon and of the black hole solutions ${\cal E}^A$ and ${\cal
E}^{B.H.}$ but could be taken arbitrarily close to $\Sigma_c^{B.H.}$ in the
limit
$\eta \to 0$. In this way the existence of a classical Euclidean motion
from
$\Sigma_c^A$ to $\Sigma_c^{\prime B.H.}$ was self evident and the
computation of
the tunneling Eq (31) was straightforward by identifying in the limit
$\Sigma_c^{\prime B.H.}$ to $\Sigma_c^{ B.H.}$. It must be shown however
that
the limit is smooth enough so that the momenta that flows between
$\Sigma_c^A$
and  $\Sigma_c^{\prime B.H.}$ in a synchronous system vanishes indeed on
the
latter hypersurface  when $\eta \to 0$. This is proven below.

Let us take, in the vicinity of the black
hole throat a Euclidean Kruskal coordinate system
$$ ds_e^2={dT_e^2+dX^2\over F^{\prime 2}(\xi)} +r^2(\xi)\,(d\theta^2+\sin^2
\theta\, d\phi^2) \eqno (A.1) $$
where
$$ F(\xi)=\sqrt{X^2+T_e^2}. \eqno(A.2) $$
Here $\xi$ is defined by (19) so that the Euclidean static coordinate
system
can be written as
$$ ds_e^2 = g_{00}(r)\,dt_e^2 + d\xi^2 + r^2(d\theta^2 +
\sin^2\theta \, d\phi^2), \eqno (A.3) $$
where
$$ {2\pi {\cal T} F(\xi)\over F^\prime (\xi)}= g_{00}^{1\over2}(\xi) \eqno
(A.4) $$

Consider a hypersurface $\Sigma (T_e^0)$ which coincide
with the $T_e = T_e^0$ hypersurface in an interval $\vert X \vert <
X_{max}$
where   $X_{max}$ is determined by the solution of
$$F^{\prime -1}(\xi) = g_{00}^{1/2}(\xi). \eqno (A.5) $$
Thus at $\vert X_{max}\vert$  the lapse functions of static
and Kruskal systems are equal. We then complete the hypersurface  $\Sigma
(T_e^0)$  by matching at $\vert X_{max}\vert$  hypersurfaces of constant
$t_e$.

For every $T_e^0>0$, we have $r>2m$ and hence one can always find a
positive
$\eta$ such that the hypersurface  $\Sigma (T_e^0) $
lays in the intersection of  ${\cal E}^A$ and ${\cal E}^{B.H.}$. We may
thus
choose  $\Sigma_c^{\prime B.H.}$ to coincide with  $\Sigma (T_e^0) $. We
must
then show that the momentum flow through  $\Sigma (T_e^0) $ vanishes in the
limit $T_e^0 \to 0$.  More precisely, defining
$$ I \equiv \int_{\Sigma (T_e^0)} \Pi^{ij} g_{ij} \, d^3x ,  \eqno
(A.6) $$
we must prove that
$$\lim_{T_e^0 \to 0} I = 0 \eqno (A.7) $$
as $I$  is indeed the surface integral in (28)
which has to vanish in order to validate the tunneling result (30).

The integral $I$ receives only contributions from the region
$-X_{max}<X<X_{max}$ and using
$$ \Pi^{ij}= {\sqrt{ g^{(3)}}\over
32\pi N}[g^{im}g^{jn}-g^{ij}g^{mn}]\partial_\tau g_{mn} \eqno(A.8)$$
where $N$ is the lapse function, one has
$$ I= - \int_0^{X_{max}} F^\prime(\xi) \left[ {\partial \over \partial
T_e}\left( {r^2(\xi) \over F^\prime (\xi)} \right) \right]\,dX  \eqno (A.9)
$$
Using (A.4) and (A.5), one gets
$$\vert I \vert = T_e^0 \int_{T_e^0}^{1 \over 2\pi {\cal T}} {1 \over
\sqrt{ F^2_e-(T_e^0)^2}}\left[ {\partial \over \partial \xi}\left(
{r^2(\xi)
\over F^\prime(\xi)}\right) \right]\,  dF  < A T_e^0 \int_{T_e^0}^{1 \over
2\pi
{\cal T}} {1 \over
\sqrt{ F^2-(T_e^0)^2}}\, dF \eqno (A.10)$$
where $A$ is a positive number. Therefore, as $T_e^0 \to 0$,
$$\vert I \vert < -A T_e^0 \ln T_e^0 \eqno (A.11) $$
and Eq (A.7) follows.
\vfil
\eject

\centerline{{\fourteenpoint{\bf{References}}}}

\item{[1]}
F. Englert,``From Quantum Correlations to Time and Entropy'' in ``The
Gardener
of Eden'' Physicalia Magazine (special issue in honour of R. Brout's
birthday),
(1990) Belgium, Ed. by P. Nicoletopoulos and J. Orloff.

\item{[2]}
A. Casher and F.Englert, Class. Quantum Grav. {\bf 9} (1992) 2231.

\item{[3]}
B. De Witt, Phys. Rev. {\bf 160} (1967) 113.

\item{[4]}
T. Banks, Nucl. Phys. {\bf 249} (1985) 332, \hfill \break
R. Brout, Foundations of Physics {\bf 17} (1987) 603, \hfill \break
R. Brout, G. Horwitz and D. Weil, Phys. Lett. {\bf B192} (1987) 318.

\item{[5]}
W. Unruh and W.H. Zurek, Phys. Rev. {\bf D40} (1989) 1064, \hfill \break
J.J. Halliwell, Phys. Rev. {\bf D39} (1989) 2912.

\item{[6]}
G. Gibbons and S. Hawking, Phys. Rev. {\bf D15} (1977) 2738; 2752.

\item{[7]}
Y. Aharonov, A. Casher and S. Nussinov, Phys. Lett. {\bf B191} (1987) 51.

\item{[8]}
J.A. Harvey and A. Strominger, ``Quantum Aspects of Black Holes", Preprint
EFI-92-41; hep-th/9209055.

\item{[9]}
A. Casher and F. Englert, Phys. Lett. {\bf B104} (1981) 117.

\item{[10]}
J. Frauendiener, C. Hoenselaers and W. Conrad,  Class. Quantum Grav. {\bf
7}
(1990) 585.

\item{[11]}
S.W. Hawking and G.F.R. Ellis, ``The Large Scale of Space-Time", (1973),
(Cambridge University Press, Cambridge, England).

\item{[12]}
P.R. Brady, J Louko and E. Poisson, Phys. Rev. {\bf D44} (1991) 1891.

\item{[13]}
J.M. Bardeen, B. Carter and S.W. Hawking, Comm. Math. Phys. {\bf 31} (1973)
161.
\item{[14]}
T.Banks,M.O'Loughlin and A.Strominger , RU-92-40 hep-th/9211030.

\vfil
\eject

FIGURE CAPTIONS.
\medskip

Figure 1. Tunneling of a nonrelativistic ``clock".

\noindent
The energy of the clock $E_c$ is represented by the dashed line. On the
left of
the turning point $a$ the clock is well represented by a forward wave
only
depicted here by a single arrow. On the right of the turning point $b$ the
amplification of the forward wave and the large concomitant backward wave
are
indicated.

 \bigskip
 \bigskip

\noindent
Figure 2. The Kruskal representation of a black hole, eventually surrounded
by static matter.

\noindent
The dashed straight line is the Euclidean axis $T_e$. The dashed circle is
the analytic continuation in Euclidean time of the solid hyperbolae
representing
trajectories $r=$constant in the static patches I and III. These are
separated
from the dynamical regions  II and IV by the horizons $r=2m_0$ where lay
the
past and future singularities $r=0$ depicted by the dashed hyperbolae. The
Schwartzschild time $t$ run on opposite directions on the two hyperbolae
$r=$constant and the Euclidean time $t_e$ spans the period ${\cal T^{-1}}$
on the
analytically continued circle.

\bigskip
\bigskip

\noindent
Figure 3. Black hole-achronon tunneling.

\noindent
The figure represents the Euclidean sections of an achronon and of its
corresponding black hole. Each point is a 2-sphere and the circles span the
Euclidean time $t_e$. The achronon geometry is depicted by thick lines and
the black hole geometry by thin lines in the region where it differs from
the
first. The picture is not on scale as there are no 3-dimensional Euclidean
imbedding of these surfaces.

\noindent
The curve $a$ is the turning hypersurface $\Sigma_c^A$.
The curve $b$ is the turning hypersurface $\Sigma_c^{B.H.}$
The curve $c$ is the hypersurface $\Sigma_c^{\prime B.H.}$ which lay in the
intersection of ${\cal E}^{B.H.}$ and ${\cal E}^A$ and tends  to
$\Sigma_c^{B.H.}$ in the limit $\eta \to 0$.

\vfil
\eject

\bye